# Phonon-laser sensing in a hetero optomechanical crystal cavity

Kaiyu Cui[1,2†*], Zhilei Huang[1,2†], Qiancheng Xu[1,2], Fei Pan[1,2], Jian Xiong[1,2], Xue Feng[1,2], Fang Liu[1,2], Wei Zhang[1,2,3], and Yidong Huang[1,2,3]

[†]These authors contributed equally to this work

[*] Corresponding author: kaiyucui@tsinghua.edu.cn

[1]Department of Electronic Engineering, Tsinghua University, Beijing 100084, China

[2]Beijing National Research Center for Information Science and Technology, Beijing, China

[3]Beijing Academy of Quantum Information Science, Beijing, China

## Abstract

Micro- and nanomechanical resonators have emerged as promising platforms for sensing a broad range of physical properties such as mass, force, torque, magnetic field, and acceleration. The sensing performance relies critically on the motional mass, the mechanical frequency, and the linewidth of the mechanical resonator. Here, we demonstrate a hetero optomechanical crystal (OMC) cavity based on a silicon nanobeam structure. The cavity supports phonon lasing in a fundamental mechanical mode with a frequency of 5.91 GHz, an effective mass of 116 fg, and a mechanical linewidth narrowing from 3.3 MHz to 5.2 kHz, while the optomechanical coupling rate of is as high as 1.9 MHz. With this phonon laser, the on-chip sensing with a resolution of $\delta\lambda/\lambda = 1.0\times10^{-8}$ can be attained, which is at least two orders of magnitude larger than that obtained with conventional silicon-based sensors. The use of a silicon-based hetero OMC cavity that harnesses phonon lasing could pave the way towards exciting, high-precision sensors that lend themselves to silicon monolithic integration and offer unprecedented sensitivity for broad physical sensing applications.

Micro- and nanomechanical resonators, allowing the detection of mass[1-4], force[5-7], torque[8], magnetic field[9,10], and acceleration[11] with ultrahigh sensitivity and over a large dynamic range, can be widely used for spectrometry, chemical analysis, biomedical diagnosis, and in consumer electronics. In particular, it is possible to drive and readout the mechanical motion via optical-only operation in optomechanical systems, creating possibilities for monolithic integration. Among the variety of optomechanical candidates that includes whispering gallery mode cavities[12-14] and membrane resonators[15], optomechanical crystal (OMC) cavities[16] are excellent candidates for use in sensing. Because OMC cavities enable strong optomechanical interaction (~ 100 kHz) with high mechanical frequency (~ GHz) and low motional mass (~ 100 fg)[17,18]. A low motional mass benefits the detection of minute masses[19-21], a high mechanical frequency results in more resistance to environmental disturbances[17,20,21] and a larger sensing bandwidth[11]. For sensitivity in sensing, it is capable of detecting mechanical motion with an imprecision at or below the standard quantum limit[5,7,22]. Unfortunately, this generally tends to be unattainable in the ambient environment of a practical setting with significantly increased mechanical dissipation[6,23-25]. This is because sensing mechanisms that rely on the detection of the mechanical resonance shift depend critically on the mechanical linewidth or mechanical $Q$ factor[19]. Efforts have been made to effectively reduce the damping loss from mechanical vibrations and thus to increase the mechanical $Q$ factor to be approximately the order of $10^6$ through the use of vacuum[11] or cryogenic measurements[18]. But vacuum or cryogenic conditions pose substantial technical challenges and limit usefulness for practical applications[26].

In an optomechanical cavity, light can interact with the mechanical motion and affect the effective mechanical damping rate[17]. When the pump light is blue-detuned and its power increased, the effective mechanical damping can be reduced to below zero, and even coherently self-sustained oscillation, that is, phonon lasing, can occur[27]. Thus by applying dissipative feedback[5,6] based on the radiation pressure of the blue-detuned pump light, the constraint of mechanical damping on the mechanical linewidth in the ambient environment can be partially relaxed[28]. Upon further increasing the power of the pump light, if the mechanical motion is excited coherently above the threshold of regenerative oscillation, highly coherent optomechanical oscillation with a narrow mechanical linewidth can be achieved. Recently, such oscillation has been induced in a silica microsphere for optomechanical spring sensing of single molecules[3]. Despite these advances, harnessing pronounced phonon laser for sensing applications has remained elusive thus far. Integrating this optomechanical sensing capability into silicon would enable high-precision sensing in a system that can be integrated with other multifunctional on-chip devices.

In this work, we demonstrate phonon-laser sensing with a hetero OMC cavity in silicon. The hetero structure, which consists of two periodic structures that separately confine the optical and mechanical defect-modes in a nanobeam cavity. By our hetero structure approach, we obtained a mechanical frequency of 5.91 GHz and a motional mass of 116 fg when the optical mode was maintained at 1573.5 nm. Using this device, the optomechanical coupling rate of the cavity was measured to be a record-high value of 1.9 MHz, and phonon lasing was achieved with a coupled optical input power of 31 μW. The mechanical damping loss was dramatically reduced with the linewidth narrowing from 3.3 MHz to 5.2 kHz after phonon lasing, corresponding to the effective mechanical $Q$ factor boosting from $1.8 \times 10^3$ to $1.1 \times 10^6$, an enhancement of three orders of magnitude. Accordingly, ultra-precise sensing with a resolution of $\delta\lambda/\lambda = 1.0\times10^{-8}$ can be achieved with this phonon laser.

**Results**

**Hetero OMC cavity**

The hetero OMC, which consists of two periodic structures (P-I and P-II) was designed to break the constraint between the design of optical and mechanical properties in an identical periodic structure of conventional OMC. By concatenating P-I and P-II with a defect region in a one-dimensional silicon nanobeam, a hetero OMC cavity was formed (Fig. 1(a)). The oblique view of a unit cell of the periodic structure, which was a silicon block with a circular air hole, is shown in the inset of Fig. 1(a). The unit cell can be determined by four geometric parameters – height ($h$), width ($w$), radius of the air hole ($r$), and length ($d$), which corresponds to the pitch of the periodic structure. P-I and P-II both consisted of 5 unit cells and possessed an $h$ value of 220 nm and a $w$ value of 460 nm, while $r$ was 136 nm for P-I and 88 nm for P-II, while $d$ was 493 nm for P-I and 400 nm for P-II. Due to the difference in geometry, the P-I and P-II structures formed the hetero OMC.

In the defect region, the radius of air holes and the spacing between holes decreased linearly insteps of 10 nm and 40 nm, respectively, from the side to center. The profiles of the optical and mechanical modes of the cavity are shown in Fig. 1(b) and (c), respectively. The mechanical frequency of the cavity reached 5.91 GHz while maintaining the optical mode at 1573.5 nm, corresponding to a frequency of 190.5 THz. Based on the mechanical mode profile, the motional mass[17] for the confined mechanical mode was calculated to be 116 fg.

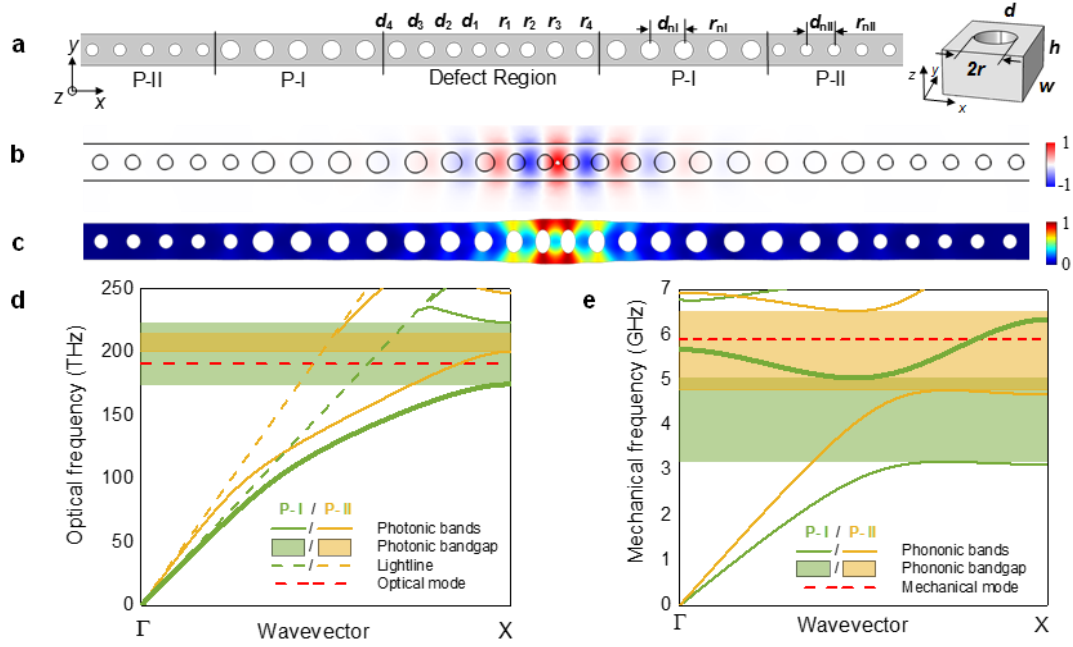

**Figure 1. a**, Top-view schematic of the hetero OMC cavity; the inset shows an oblique-view schematic of the unit cell that forms the periodic structure. The unit cell can be determined by four geometric parameters – height ($h$), width ($w$), radius of the air hole ($r$), and length ($d$), which corresponds to the pitch of the periodic structure. **b, c,** Mode profiles of the electric y-component of the optical mode (**b**) and that of the displacement of the mechanical mode (**c**). **d, e,** TE optical bands and y- and z-symmetric mechanical bands of the P-I (green) and P-II (yellow) structures, respectively. The red dashed lines in **d** and **e** represent the frequency of the optical and mechanical modes, respectively. Due to the unequal period of the P-I and P-II structures, the X points correspond to different wavevectors, and the lightlines of the P-I (yellow dashed line) and P-II (green dashed line) structures do not overlap in the photonic band diagram.

The confinement mechanism for the optical and mechanical modes in the hetero structure can be analyzed using the photonic and phononic bands formed by the two periodic structures. Figure 1(d) and (e) show the TE photonic and $y$-, $z$-symmetric phononic band diagrams of these two periodic structures. Due to the difference in geometry, the P-I and P-II structures possessed different photonic and phononic bands. The green color represents the P-I structure, and the yellow color represents the P-II structure. Since the period and hole radius of P-II were smaller than those of P-I, the central frequency of the photonic bandgap of P-II was higher, while the range was smaller, as shown in Fig. 1(d). The smaller period and hole radius of the P-II structure also contributed to a higher frequency for the mechanical bandgap, as shown in Fig. 1(e).

The defect modes are depicted by the red dashed lines in the band diagrams shown in Fig. 1(d) and (e). The optical defect mode was located inside the photonic bandgap of the P-I structure. However, the shrinkage of both the radius of the air holes and the spacing between holes contributed to a higher-frequency mechanical defect mode, which exceeded the phononic bandgap of the P-I structure. Nevertheless, since the frequency of the mechanical defect mode was located inside the mechanical bandgap of the P-II structure, the whole cavity confined the mechanical mode by the hetero OMC. Therefore, the optical and mechanical modes were confined separately by the P-I and P-II structures using the hetero OMC cavity.

**Fabrication and measurement of the hetero OMC cavity**

To verify the degree and effectiveness of the mode confinement in the hetero structure, we fabricated three different cavity structures for comparison: i) a hetero OMC cavity, ii) an optomechanical cavity without a hetero structure, and iii) an optomechanical cavity with an acoustic radiation shield. The acoustic radiation shield[29] was a two-dimensional cross periodic structure capable of providing a complete phononic bandgap between 5.16 and 6.41 GHz. Top-view SEM images of the three different structures are given in Figs. 2a-c. They were probed by a tapered fiber in an ambient environment, and their optical and mechanical spectra are shown in Figs. 2(d) and (e). The optical spectra show that all three cavity structure types possessed similar optical and mechanical frequencies as well as optical $Q$-factors with the order of $10^4$. However, the mechanical $Q$-factor of the OMC cavity without a hetero structure was $2.6 \times 10^2$, which is much smaller than the values of the hetero OMC cavity and the OMC cavity with a radiation shield, both of which showed a mechanical *Q*-factor of approximately $1.8\times10^3$. This indicates that the acoustic waves leaking through the P-I structure were blocked by the outer periodic structures of the hetero OMC cavity and by the OMC cavity with an acoustic radiation shield, and P-II structure of the hetero OMC delivered a confinement effect similar to that of the acoustic radiation shield in an ambient environment. As a result, the optical and mechanical modes were successfully confined separately by the two periodic structures in the hetero OMC cavity and by the OMC cavity with an acoustic radiation shield, which has been widely used for OMCs[29,30]. Compared with the two-dimensional acoustic radiation shield, the hetero OMC cavity designed with one-dimensional nanobeam structure is more suitable for use in integrated circuits.

In addition, thanks to the design flexibility offered by the separate confinement of the optical and mechanical modes in the hetero structure, we were able to demonstrate a strong optomechanical coupling. The measured optomechanical coupling rate ($g_0/2\pi$) of the cavity was 1.9 MHz, which is the highest rate among reported optomechanical systems (see Supplementary Information, section S1).

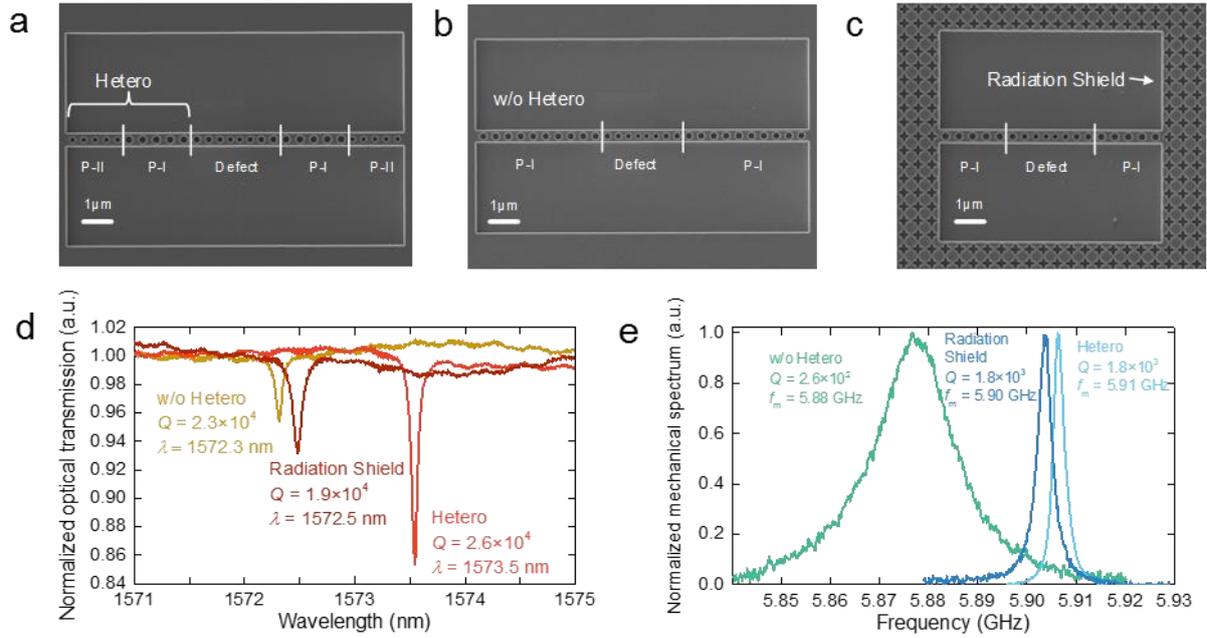

**Figure 2. a-c**, Top-view SEM images of the hetero OMC cavity (**a**), OMC cavity without a hetero structure (**b**), and OMC cavity with radiation shield (**c**). **d**, **e**, Optical (**d**) and mechanical (**e**) spectra of the three types of OMC cavities.

**Phonon laser sensing in the hetero OMC cavity**

Due to the good confinement of the mechanical mode and the large optomechanical coupling rate, phonon lasing was demonstrated in the hetero OMC cavity with low coupled optical power. Figure 3(a) shows the normalized phonon number of the hetero OMC as a function of the coupled optical power when the pump light is blue-detuned. Here, the phonon number is normalized to that excited by the thermal environment, which is $1.06 \times 10^3$ at room temperature (see Supplementary Information, section S2). The measured threshold of the coupled optical power was 31 μW. Beyond this threshold, the mechanical linewidth was dramatically suppressed, reducing from 3.3 MHz to 5.2 kHz, as shown in Fig. 3(b), which corresponds to an effective mechanical $Q$-factor boosting from $1.8 \times 10^3$ to $1.1 \times 10^6$.

This suggests that the sensing resolution of sensing mechanisms that rely on detecting the frequency shift of the mechanical resonator can be greatly improved by the coherently-narrowed mechanical linewidth that arises after phonon lasing. Since the mechanical frequency of the phonon laser will shift with detuning between the pump light and the optical resonance due to the optomechanical spring effect[3,17], the coherently-enhanced sensing resolution can be analyzed as follows.

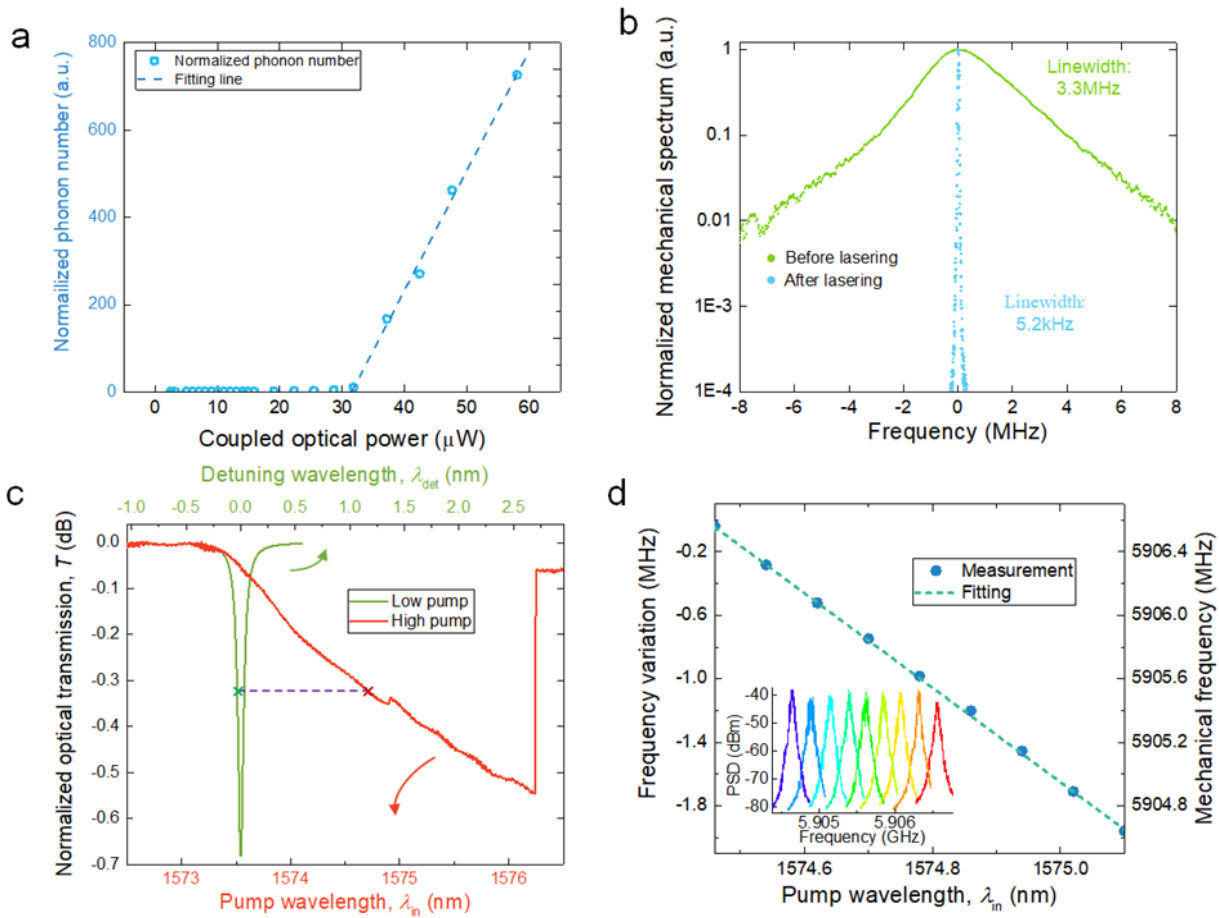

**Figure 3. a**, Normalized phonon number of the hetero OMC cavity as a function of coupled optical power. **b**, Normalized mechanical spectra of the hetero OMC cavity before and after phonon lasing. **c**, Normalized optical transmission under low and high pump power as a function

of the pump wavelength and the detuning wavelength. The two crosses joined by the purple dashed line indicate the same detuning. Here, the ratio between the optical detuning and optical decay rate was 0.50 when the linewidth of the mechanical spectrum achieved a minimum. **d**, Measured mechanical frequency as a function of the pump wavelength and the fitting line. The inset shows the shift in the mechanical peak under lasing at different pump wavelengths.

We now discuss the mechanism by which our optomechanical cavity can be used to perform coherently enhanced sensing. When there is a small variation in the refractive index in the cavity's surroundings, such as a new particle that appears near the hetero cavity, the optical resonance of the cavity will shift and cause detuning between the fixed laser input and the optical resonance. Because the mechanical mode is sensitive to the laser-cavity detuning that is induced by the optical spring effect[3,17], the refractive index variation can be determined by measuring the mechanical frequency shift instead of the shift in optical mode used in conventional methods[2,31]. However, in addition to the index variation induced by the detected particle, the optical frequency of the cavity will be redshifted due to heat deposition when the OMC cavity is pumped with relatively high power to realize phonon lasing, as shown in Fig. 3(c). Thus, with a view to phonon laser sensing, two effects will lead to the resonance shift of the optical cavity and contribute to laser-cavity detuning. The first effect is the variation in the refractive index caused by a detected particle. The second is changes in the index due to the heat effect under high pump power. The wavelength variation caused by the former effect – that is, the laser-cavity detuning not attributable to influence of heat under high pump power – needs to be determined for the purposes of practical sensing. It can be deduced as follows.

Since the same optical transmission under different pumping powers corresponds to the same level of detuning for an optical cavity[3], the two crosses linked by the purple dashed line in Fig. 3 (c) indicate an equivalent detuning. Thus the ratio between the laser-cavity detuning not attributable to the influence of heat and the mechanical frequency variation $d\lambda_{det}/df_m$ can be expressed as follows:

$$\frac{d\lambda_{det}}{df_m} = \frac{d\lambda_{det}}{dT} \times \frac{dT}{d\lambda_{in}} \times \frac{d\lambda_{in}}{df_m} \qquad (1)$$

Here $\lambda_{det}$ is the detuning wavelength between the laser and the cavity resonance without the heat effect, which corresponds to the cavity resonance-shift caused by the small variation in refractive index induced by a particle; $T$ is the optical transmission power; $\lambda_{in}$ is the pump-light wavelength; and $f_m$ is the mechanical frequency. In our experiments, $dT/d\lambda_{det}$ and $dT/d\lambda_{in}$ can be deduced from the slope of the optical transmission in Fig. 3(c). Their values were measured as 20.9 dB/nm and -0.195 dB/nm, respectively. A value of $d\lambda_{in}/df_m = 3.36 \times 10^{-4}$ nm/kHz was obtained from the measured mechanical frequency as a function of the pump wavelength in Fig. 3 (d). Here, the mechanical peaks obtained with lasing under different pump wavelengths are also presented in the inset of Fig. 3 (d). Based on these parameters, $d\lambda_{det}/df_m$ was calculated as $-3.1 \times 10^{-6}$ nm/kHz via equation (1) for the demonstrated hetero cavity.

As the mechanical spectrum is dramatically narrowed after phonon lasing, the detectable mechanical frequency shift $\delta f_m$ could be minimized to 5.2 kHz, yielding a wavelength resolution $\delta\lambda = d\lambda_{det}/df_m \times \delta f_m$ of $1.6 \times 10^{-5}$ nm. Therefore, a detection precision of $\delta\lambda/\lambda = 1.0 \times 10^{-8}$ can be achieved with $\lambda = 1572.3$ nm, which represents an enhancement of three orders of magnitude compared with the mechanical linewidth of 3.3 MHz before phonon lasing.

## Discussion and conclusions

In this study, we have demonstrated that the use of hetero OMC cavities can separately confine the optical and mechanical modes using two types of periodic structures. Because the optical and mechanical modes are no longer confined by an identical periodic structure in conventional OMCs, the hetero structure led to a high mechanical frequency of 5.91 GHz while maintaining the optical mode at 1573.5 nm. The mechanical $Q$-factor of the hetero OMC was found to be $1.8\times10^3$, which is much higher than that of OMCs without a hetero structure ($2.6\times10^2$), and is close to that offered by OMCs with an acoustic radiation shield, thus indicating that the optical and mechanical modes were well confined by the two periodic structures. The separate confinement mechanism introduced by the hetero structure is also responsible for boosting the coupling rate between the optical and mechanical modes, which was found to be as high as 1.9 MHz in the hetero cavity.

With a hetero OMC cavity implemented in silicon, pronounced phonon lasing has been shown to be viable for high-precision sensing applications. Due to the coherence-related enhancement offered by the phonon lasing, mechanical damping losses were dramatically reduced, with linewidth narrowing from 3.3 MHz to 5.2 kHz in an ambient environment. This corresponds to ultra-precise sensing with $\delta\lambda/\lambda = 1.0\times10^{-8}$. To date, the largest optical $Q$ factor obtained for silicon-based microcavities is $10^6$ [ref. 25], yielding a sensing resolution of $\delta\lambda/\lambda \sim 10^{-6}$. We have thus demonstrated an improvement in the sensing resolution of two orders of magnitude.

By coupling the hetero OMC cavity with an integrated evanescent coupling waveguide, the cavity could be readily used as a functional component and integrated with other on-chip devices for use in practical applications. Moreover, the mode confinement approach adopted here could be applied to two-dimensional cavities[32] or waveguides[33] to further improve their mechanical properties.

The use of a silicon-based hetero OMC cavity that harnesses phonon lasing in an ambient environment offers a potentially powerful solution for integrated, on-chip sensing with a resolution two orders of magnitude larger than that obtained with conventional silicon-based sensors.

## Methods

### Simulations

The photonic band structures in this paper were calculated using the plane wave expansion (PWE) method. Other simulations, including those of the phononic bands and the optical and mechanical modes, were performed using the finite element method (FEM). To calculate the optomechanical coupling rate, the optical and mechanical modes were simultaneously solved in the same mesh system so that the domain and surface integration for calculation of the photoelastic effect and moving boundary effect[29] could be computed.

### Fabrication

All patterned structures were first defined by electron beam lithography (EBL) and transferred to the device layer of silicon-on-insulator (SOI) chips by inductively coupled plasma (ICP) etching. For the tapered-fiber-coupled cavities, the dry-etched structures were directly wet-etched by buffered hydrofluoric acid (BHF) to form suspended structures before measurement. For the cavity integrated with an evanescent waveguide, a 600-nm-thick silica layer was deposited via plasma-enhanced chemical vapor deposition (PECVD) after ICP etching, and the chips were back-ground to a thickness of 100 μm thick for future cleavage. After that, 13-μm-width strip windows above the cavities were formed by aligned ultraviolet lithography. The buried and deposited silicon oxide layers above and below the cavities were removed by wet-etching with BHF to form suspended structures. Finally, the chips were cleaved and prepared for testing.

**Measurements**

A tunable laser light source was used. The structures without coupling waveguides were optically coupled using a tapered fiber, and the coupling waveguide of the integrated structure was end-fire-coupled by a lensed fiber. For both measurement schemes, the output was split into two channels. One port was connected to a low-frequency optical power monitor, from which the optical spectrum was obtained by sweeping the wavelength of the laser. The other port was connected to a high-frequency (12.5-GHz) optical receiver, and its output was connected to an electrical spectrum analyzer to obtain the mechanical spectra. The optomechanical coupling rate of the cavity was measured with the assistance of an electro-optic modulator for calibration.

**Code availability**

Custom code used in this study is available from the corresponding authors upon reasonable request.

## Data availability

The data that support the plots within this paper and other findings are available from the corresponding author upon reasonable request.

## Acknowledgements

The authors would also like to thank Dr. Di Qu and Mr. Guoren Bai of Innovation Center of Advanced Optoelectronic Chip and Institute for Electronics and Information Technology in Tianjin, Tsinghua University for their help with device fabrication. This work was supported by the National Key R&D Program of China under Contracts No. 2017 YFA0303700, the National Natural Science Foundation of China (Grant No. 61775115, 91750206, 61575102, and 61621064), the Opened Fund of the State Key Laboratory on Integrated Optoelectronics (No. IOSKL2016KF01), and Beijing Innovation Center for Future Chips, Tsinghua University.

## Author Contributions

K.C. and Z.H. conceived the study. K.C. performed the theoretical analysis, Z.H. conducted the fabrication and measurement, and Q.X., F.P., and J.X. analyzed the results. K.C. and Z.H. wrote the paper. X.F., F.L., W.Z. and Y.H. discussed the results and reviewed the manuscript.

# Supplementary Information

# Phonon laser sensing in a hetero optomechanical crystal cavity

## S1. Measuring optomechanical coupling rate ($g_0$)

We measure the optomechanical coupling rate with the method similar with that proposed by M. L. Gorodetksy *et al.* and K. Balram *et al.*[1,2], and that used by Z. Huang *et al.*[3] The overall idea is as follows. The power of the optical signal detected by the photoreceiver around the oscillation frequency is related with the mechanical vibration intensity with $g_0$. This power can be measured quantitatively by the electronic spectrum analyzer with the assistant of the intensity electro-optic modulator to calibrate. Thus, the optomechanical coupling rate ($g_0$) can be obtained.

The evolution equations describing the cavity optomechanical system are[4]

$$\dot{a} = -\frac{\kappa}{2}a + i(\Delta + Gx)a + \sqrt{\kappa_{ex}}\, s_{in}, \tag{S1}$$

$$\ddot{x} = -m\Omega_m^2 x - m\Gamma\dot{x} + \hbar G|a|^2, \tag{S2}$$

In these equations, $a$ is the optical amplitude in the cavity. $\Delta$ is the laser detuning from the cavity resonance. $\kappa$ and $\kappa_{ex}$ are the total decay rate of the optical mode and the optical coupling rate between the optical input channel and the cavity, respectively. $\Gamma$, $x$, $m$, and $\Omega_m$ denote the decay rate, the displacement, the effective mass, and the intrinsic angular frequency of the mechanical mode, respectively. $G$ is the optomechanical coupling rate divided by the zero point fluctuation displacement, i.e. $g_0/x_{zpf}$, representing the optical frequency shift per displacement. $s_{in}$ is the amplitude of the pump light.

Assume the mechanical vibration to be $x(t) = x_0 \cos(\Omega_m t)$, substitute it into equation (S1) and assume a stable optical input, we have the normalized optical amplitude in the cavity can be expressed as

$$a(t) = s_{in}\sqrt{\eta\kappa}\mathcal{L}(0)\left(1 - \frac{ix_0 g_0 \mathcal{L}(\Omega_m)}{2x_{zpf}}e^{-i\Omega_m t} - \frac{ix_0 g_0 \mathcal{L}(\Omega_m)}{2x_{zpf}}e^{i\Omega_m t}\right), \tag{S3}$$

where function $\mathcal{L}(\Omega)$ is defined as

$$\mathcal{L}(\Omega) = \frac{1}{-i(\Delta + \Omega) + \kappa/2} \tag{S4}$$

to simplify the derivation. The normalized output optical amplitude is

$$s_{out}(t) = s_{in} - \sqrt{\kappa_{ex}}\, a(t) \tag{S5}$$

and the relation between the output optical power ($I_{OM}$) and the normalized output optical amplitude is

$$I_{OM}(t) = \hbar\omega_o \left|s_{out}(t)\right|^2. \tag{S6}$$

The output optical power affected by the cavity optomechanical system should has the form of

$$I_{OM}(t) = I_{0OM} + I_{1OM}\cos(\Omega_m t + \theta). \tag{S7}$$

The power detected by the electrical spectrum analyzer around the oscillation frequency should be proportional to $I_{1OM}^2$, which means $P_{ESA,OM} = k_{ESA} I_{1OM}^2$. So, $I_{1OM}^2$ is the term we concern most. From equation (S6) and (S7), we have

$$I_{1OM}^2 = \frac{1024 \Delta^2 \eta^2 \kappa^2 \left((-1+\eta)^2 \kappa^2 + \Omega_m^2\right)}{\left(4\Delta^2 + \kappa^2\right)^2 \left(\left(4\Delta^2 + \kappa^2\right)^2 + 8\Omega_m^2\left(-4\Delta^2 + \kappa^2 + 2\Omega_m^2\right)\right)} \left(\frac{\hbar\omega_L x_0 g_0}{x_{zpf}}\right)^2 \left|s_{in}\right|^4, \tag{S8}$$

where $\eta = \kappa_{ex}/\kappa$. The value of equation (S8) is related with optical detuning ($\Delta$). When $\Delta = \dfrac{\sqrt{\kappa^2 + 4\Omega_m^2}}{2\sqrt{3}}$, its value will be maximum as

$$I_{1OM}^2 = \frac{27\eta^2 \kappa^2 \left((-1+\eta)^2 \kappa^2 + \Omega_m^2\right)}{\left(\kappa^2 + \Omega_m^2\right)^3} \left(\frac{\hbar\omega_L x_0 g_0}{x_{zpf}}\right)^2 \left|s_{in}\right|^4. \tag{S9}$$

During the experimental measurement, we fine-tuned the wavelength of the input laser to obtain the maximum signal in the electrical spectrum analyzer. Thus equation (S9) can be used to estimate the power detected by the electrical spectrum analyzer around oscillation frequency.

The ration between $x_0$ and $x_{zpf}$ can be obtained by $x_0^2 / x_{zpf}^2 = 4 n_m$, where $n_m$ is the phonon number of the cavity and can be obtained by the measurement of Fig. 3(a) of the main text.

To calibrate the power obtained by the electrical spectrum analyzer, we used an intensity electro-optic modulator, the method of which is the same as the work done by Z. Huang *et al.*[3] By knowing the power detected by the electrical spectrum analyzer around the oscillation frequency, we can have the optomechanical coupling rate ($g_0/2\pi$) from equation (S9), which is 1.9 MHz for the hetero optomechanical crystal cavity.

## S2. Normalized phonon number

Assuming the mechanical motion has the form of $x = x_0 \sin(\Omega_m t)$. Then, the light amplitude in the cavity can be deduced from (S1-2) and expressed as

$$a = a_0 + a_1 e^{i\omega t} + a_{-1} e^{-i\omega t}, \tag{S10}$$

where

$$a_0 = \frac{\sqrt{\kappa_{ex}}}{\frac{\kappa}{2} - i\Delta} s_{in}, \tag{S11}$$

$$a_1 = \frac{G a_0 x_0}{2} \frac{1}{-i(\Delta - \Omega_m) + \kappa/2}, \tag{S12}$$

$$a_{-1} = -\frac{G a_0 x_0}{2} \frac{1}{-i(\Delta + \Omega_m) + \kappa/2}. \tag{S13}$$

As the intensity of the output light is promotional to $\left| s_{in} - \sqrt{\kappa_{ex}} a \right|^2$, the power of the signal detected by the electrical spectrum analyser around the oscillation frequency ($P_{ESA}$) is promotional to

$$\left[ \frac{1024 \Delta^2}{\left(4\Delta^2 + \kappa^2\right)^2 \left(\kappa^2 + 4(\Delta - \Omega_m)^2\right) \left(\kappa^2 + 4(\Delta + \Omega_m)^2\right)} \left((\kappa - \kappa_{ex})^2 + \Omega_m^2\right) \kappa_{ex}^2 G^2 \right] x_0^2 P_{in}^2, \tag{S14}$$

where $P_{in}$ is the optical power of the input light. The intensity of the mechanical motion is proportional to the phonon number, i.e. $n_m \propto x_0^2$. Consequently,

$$n_m \propto P_{ESA} / P_{in}^2, \tag{S15}$$

When detected by light with low power, the phonon number of the mechanical motion is nearly that exited by the thermal environment, which means the normalized phonon number is 1.